\documentclass[10pt,twocolumn,aps,manuscript,superscriptaddress]{revtex4}
\usepackage[latin9]{inputenc}
\usepackage{amsmath}
\usepackage{amssymb}
\usepackage{graphicx}
\usepackage{subscript}
\usepackage[unicode=true,
 bookmarks=true,bookmarksnumbered=false,bookmarksopen=false,
 breaklinks=false,pdfborder={0 0 1},backref=false,colorlinks=false]
 {hyperref}
\hypersetup{pdftitle={Test of Lorentz Invariance and of space-time fluctuations with a crossed-cavities laser experiment}}

\makeatletter
\@ifundefined{textcolor}{}
{%
 \definecolor{BLACK}{gray}{0}
 \definecolor{WHITE}{gray}{1}
 \definecolor{RED}{rgb}{1,0,0}
 \definecolor{GREEN}{rgb}{0,1,0}
 \definecolor{BLUE}{rgb}{0,0,1}
 \definecolor{CYAN}{cmyk}{1,0,0,0}
 \definecolor{MAGENTA}{cmyk}{0,1,0,0}
 \definecolor{YELLOW}{cmyk}{0,0,1,0}
}

\makeatother

\begin{document}

\title{Resonator with ultra-high length stability as a probe for Equivalence-Principle-violating
physics}

\author{E.~Wiens, A.Yu.~Nevsky, and S. Schiller}

\address{Institut für Experimentalphysik, Heinrich-Heine-Universtität Düsseldorf,
40225 Düsseldorf, Germany}
\begin{abstract}
In order to investigate the long-term dimensional stability of matter,
we have operated an optical resonator fabricated from crystalline
silicon at 1.5~K continuously for over one year and repeatedly compared
its resonance frequency $f_{res}$ with the frequency of a GPS-monitored
hydrogen maser. After allowing for an initial settling time, over
a 163-day interval we found a mean fractional drift magnitude $|f_{res}^{-1}df_{res}/dt|<1.4\times10^{-20}$/s.
The resonator frequency is determined by the physical length and the
speed of light, and we measure it with respect to the atomic unit
of time. Thus, the bound rules out, to first order, a hypothetical
differential effect of the universe's expansion on rulers and atomic
clocks. We also constrain a hypothetical violation of the principle
of Local Position Invariance for resonator-based clocks and derive
bounds for the strength of space-time fluctuations.   
\end{abstract}
\maketitle
In this paper we address experimentally the question about the intrinsic
time-stability of the length of a macroscopic solid body. This question
is related to the question about time-variation of the fundamental
constants and effects of the expansion of the universe on local experiments.
It may be hypothesized that, in violation of the Einstein Equivalence
Principle (EP), the expansion affects the length of a block of solid
matter and atomic energies to a different degree. The length, defined
by a multiple of an interatomic spacing, can be measured by clocking
the propagation time of an electromagnetic wave across it. This procedure
effectively implements the Einstein light clock, or, in modern parlance,
an electromagnetic resonator. The hypothetical differential effect
would show up as a time-drift of the ratio of the frequency $f_{res}$
of an electromagnetic resonator and of an atomic (or molecular) transition
($f_{atomic}$). A resonator and an atom are dissimilar in the sense
that the former's resonance frequency intrinsically involves the propagation
of an electromagnetic wave, while the latter does not. Specifically,
the time-drift would violate the principle of Local Position Invariance
(LPI) of EP. The natural scale of an effect due to cosmological expansion,
here the fractional drift rate $D_{res-atomic}=$ $(f_{atomic}/f_{res})d(f_{res}/f_{atomic})/dt$,
is the Hubble constant $H_{0}\simeq2.3\times10^{-18}$/s. Extensive
work in the last decade has ruled out that an effect of this order
exists between different \emph{atomic and molecular} frequency standards
\cite{Uzan2011}.

The suitable regime in which to investigate the dimensional stability
of matter is at cryogenic temperature, when the thermal expansion
coefficient and the thermal energy content of matter are minimized.
Ideally, during the cooling down and then permanence at cryogenic
temperature, a stable energy minimum of the solid is reached. The
expected high dimensional stability and the magnitude of $H_{0}$
lead to a challenging measurement problem: how to resolve tiny length
changes, and how to suppress the influence of extrinsic disturbances.
The problem can be addressed by casting the solid matter into an electromagnetic
resonator of appropriate shape, by supporting it appropriately, and
by measuring its resonance frequency using atomic time-keeping and
frequency metrology instruments, which indeed permit ultra-high measurement
precision and accuracy. 

Cryogenically operated resonators \cite{Turneaure1983} have been
developed for microwave and optical frequencies. These represent
a viable approach for realizing oscillators having ultra-high stability
both on short and on long time scale (months). Therefore, they have
found applications for tests of EP \cite{Turneaure1983,Braxmaier2002,Muller2003a,Antonini2005,Schiller2004,Tobar2010}
and as local oscillators in connection with microwave and optical
atomic clocks \cite{Millo2009a,Kessler2012a}. 

Recent cryogenic microwave sapphire resonators exhibit nonzero fractional
drifts $(D=f^{-1}df/dt)$ of $-1.7\times10^{-18}/$s \cite{Tobar2010}
and $4\times10^{-19}/$s \cite{Hartnett2012} and the lowest value
reported was $-1.9\times10^{-20}/$s during a 9-day long interval
\cite{Hartnett2006}. For a sapphire cryogenic optical resonator \cite{Seel1997}
a mean drift smaller than $6.4\times10^{-19}/$s was measured \cite{Storz1998}.
Recently, silicon resonators (first studied in Ref.~\cite{Richard1991})
were investigated at 123~K \cite{Kessler2012a} and at 1.5~K \cite{Wiens2014,Wiens2015}.
Silicon's advantages are the availability of single crystals of large
size at affordable cost, its easy machineability, a flexibility in
the choice of resonator shape, and superpolished mirror substrates
allowing high-reflectivity mirrors. Important properties of silicon
resonators, such as thermal expansion coefficient, thermal response,
resonator linewidth and throughput have already been described in
the references given. The 123~K silicon resonator system exhibited
an average drift of less than $5\times10^{-19}$/s over an interval
of 70~days \cite{Hagemann2014}. 

In earlier work on cryogenic resonators, cryostats using liquid coolants
were often employed. With the advent of ``dry'' cryostats, first
applied to optical resonator experiments in Ref.~\cite{Antonini2005},
the operation of cryogenic resonators has become possible with reduced
maintenance effort and without the disturbances due to periodic refill
of cryogens, opening up new opportunities. Using this technology,
the present study was therefore able to investigate a silicon resonator
operated at 1.5~K continuously for 420 days. This low temperature
is attractive not only for reducing thermal expansion (and consequently
reduced requirements for active temperature stabilization) and thermally
activated processes, but also for reducing the thermal noise of the
resonator's components: spacer, mirror substrates, and mirror coatings
\cite{Wiens2014}. 

We demonstrate here that a crystalline resonator can exhibit outstanding
dimensional stability. This performance can be used to test for EP-violating
effects. It also has technological potential as an oscillator with
performance improved compared to the well-established masers. 

\begin{figure}[t]
\begin{minipage}[t]{1\columnwidth}%
\noindent \begin{center}
\includegraphics[bb=190bp 100bp 640bp 500bp,clip,width=1\columnwidth]{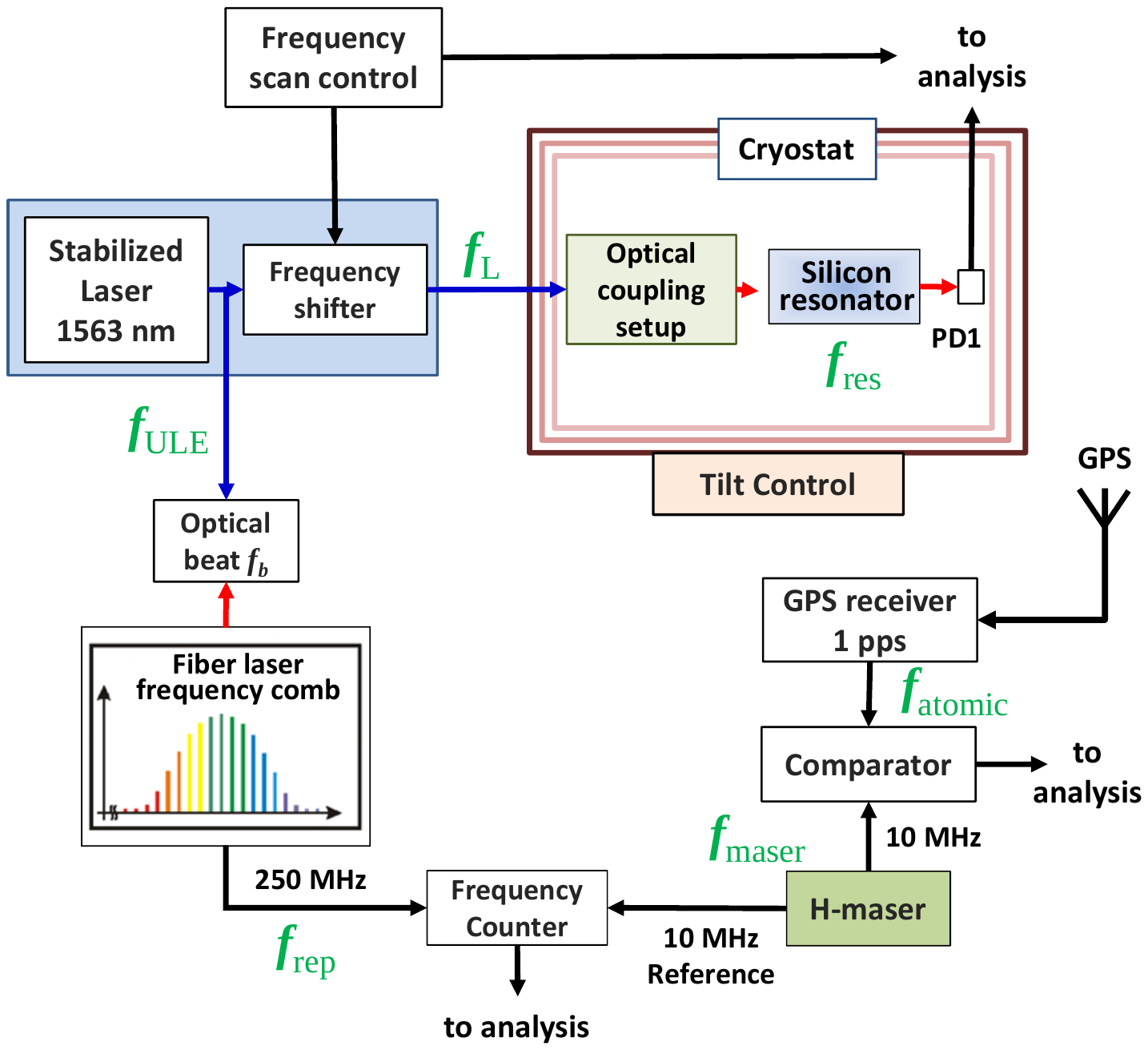}
\par\end{center}%
\end{minipage}

\protect\caption{\label{fig:Comb Lock-1}(Color online). Principle of the experiment.
The optical resonance frequency $f_{res}$ of a silicon resonator
is compared with a radiofrequency $f_{maser}$ provided by a hydrogen
maser. This is done via the intermediary of a laser whose wave of
frequency $f_{L}$ interrogates the resonator and is also measured
by an optical frequency comb (pulse repetition rate $f_{rep}$). The
maser frequency is itself compared with a 1-Hz-signal ($f_{atomic}$)
obtained from GPS satellites. }
\end{figure}

\textit{Overview of the apparatus.} Fig.~\ref{fig:Comb Lock-1} (see
also \cite{SupplementalMaterial-Cryoresonator2016-for-arxiv-tex-version})
shows the concept of the experiment, which is an extension of our
previous work \cite{Wiens2014}. A silicon resonator is operated in
a pulse-tube cooler (PTC) cryostat with Joule-Thomson (JT) cooling
stage,  providing a base temperature 1.5~K. 

The silicon resonator is 25~cm long and consists of a cylindrical
spacer and two optically contacted silicon mirror substrates \cite{SupplementalMaterial-Cryoresonator2016-for-arxiv-tex-version}.
Its linewidth is 2.0~kHz. The resonator is oriented horizontally
and supported by wires inside a copper frame \cite{SupplementalMaterial-Cryoresonator2016-for-arxiv-tex-version}.
The supports' symmetry is such that to first order the thermal expansion
or length drift of the frame does not affect the resonator length. 

In the frequency-scan interrogation technique (Fig.~\ref{fig:Comb Lock-1}),
the resonator's TEM\textsubscript{00} mode frequency $f_{res}$ is
read out by a $1.56\,\mu$m (192~THz) external-cavity semiconductor
laser (``Laser~1'' in \cite{SupplementalMaterial-Cryoresonator2016-for-arxiv-tex-version}),
which is frequency-stabilized to a room-temperature reference resonator
(frequency $f_{ULE}$), having short-time fractional frequency instability
$5\times10^{-15}$ and drift $6\times10^{-16}/{\rm s}$. During read-out,
the laser frequency is offset to $f_{L}$ and is repeatedly scanned
in time over a range of a few kHz across the silicon resonator resonance
(line center frequency $f_{res}$). We record on the photodetector
PD1 the power of the laser wave transmitted through the resonator,
fit a lineshape model to each scan's data, and extract the frequency
offset $f_{res}-f_{ULE}$ . Simultaneously, the laser frequency corresponding
to $f_{ULE}$ is measured by an erbium-fiber-laser-based femtosecond
frequency comb, using a hydrogen maser ($f_{maser}$) as reference.
From these two measurements we obtain $f_{res}$ in units of $f_{maser}$.
If $f_{maser}$ was constant in time, the long-term variation of $f_{res}$
would mainly be given by the long-term variation of the length $l$
of the silicon crystal resonator spacer and (hypothetically) of the
speed of light $c$, $\Delta f_{res}/f_{res,0}=\Delta c/c_{0}-\Delta l/l_{0}$,
where $f_{res,0}$, $l_{0}$ and $c_{0}$ are the resonance frequency,
the spacer length, and the speed of light at a reference time $t_{0}$,
respectively.

\textit{Systematic effects.} The experiment requires maintaining the
operating parameters of the system as stable as possible. Several
systematic effects were investigated. 

\begin{figure}[h]
\begin{minipage}[t]{1\columnwidth}%
\begin{center}
\includegraphics[width=0.95\columnwidth]{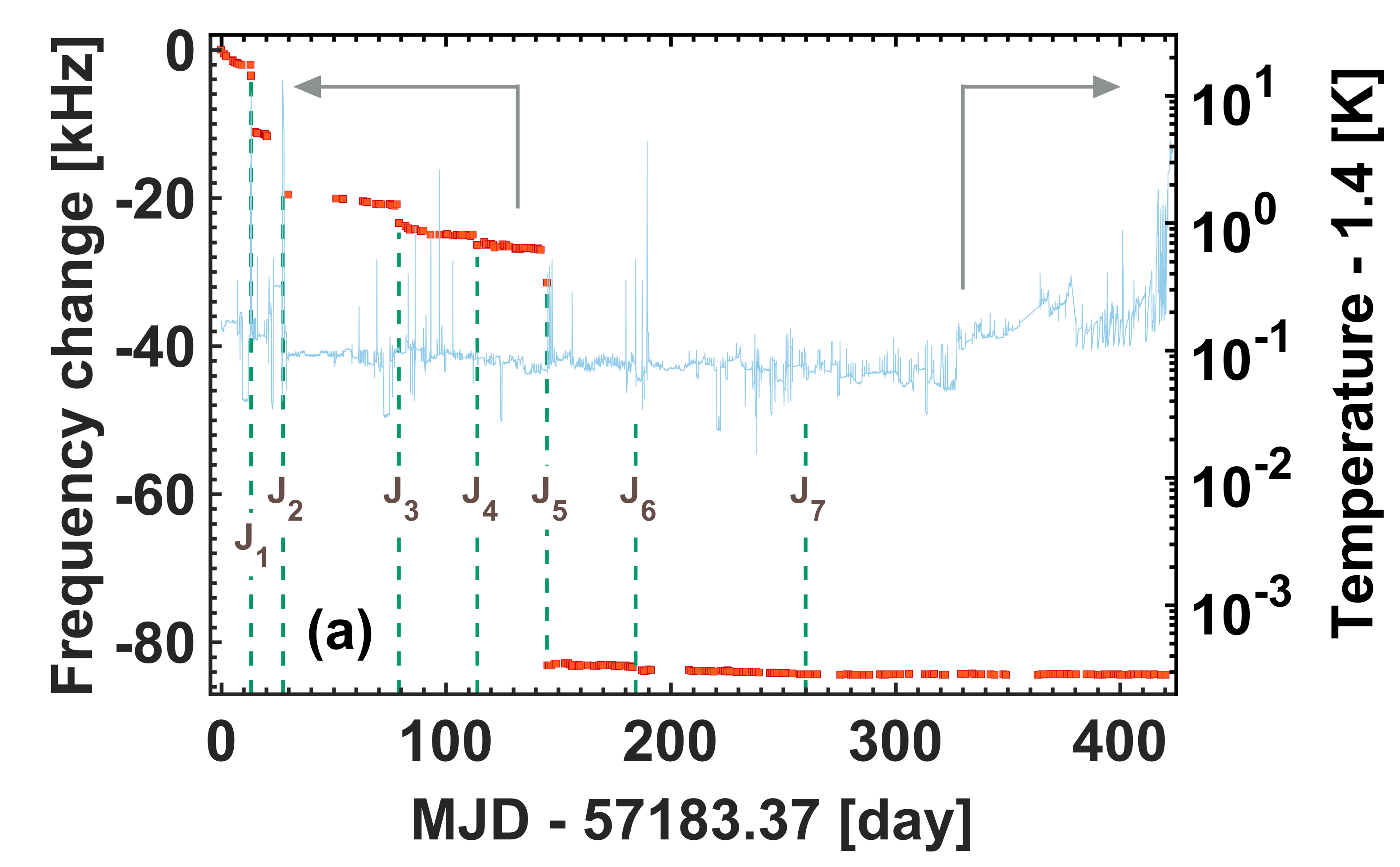}
\par\end{center}%
\end{minipage}

\begin{minipage}[t]{1\columnwidth}%
\begin{center}
\includegraphics[width=0.95\columnwidth]{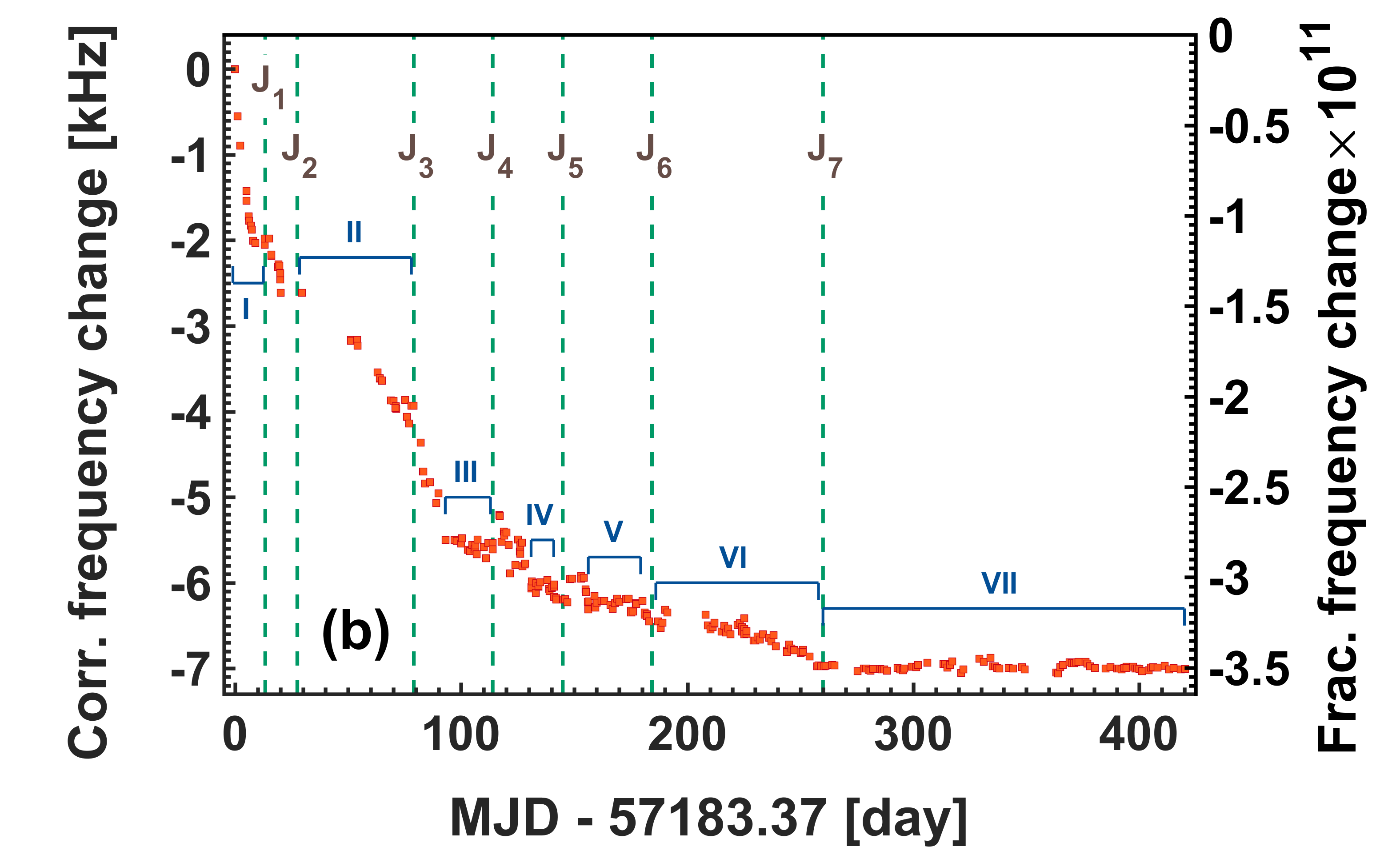}
\par\end{center}%
\end{minipage}

\protect\caption{\label{fig:Si resonator frequency}(Color online). (a): Variation
$\Delta f_{res}$ of the optical frequency of the silicon resonator
(left scale) and its temperature (right scale) over time. During intervals
I~ -~VI, the Pound-Drever-Hall locking technique was used. The shown
data of interval VII was obtained using the frequency scan technique.
Time zero corresponds to the first measurement, performed four days
after reaching base temperature (1.5~K). (b): Corrected Si resonator
frequency change, obtained by removing frequency jumps that occurred
at $J_{1},\ldots,\, J_{6}$.}
\end{figure}

(1) Temperature: The cryostat was operated at its base temperature
and no active temperature stabilization of the resonator was used,
since it was unnecessary in the present context. Fig.~\ref{fig:Si resonator frequency}
shows the temperature over a period of approximately 420~days with
a typical peak-peak variation of 30~mK. The thermal sensitivity of
the resonator being $|f_{res}^{-1}df_{res}/dT|<1\times10^{-12}/$K
at 1.5~K \cite{Wiens2014,Wiens2015}, this corresponds to a peak-peak
fractional frequency deviation $<\,4\times10^{-15}$, which is not
of importance here. On day 327 the operating temperature had to be
raised to 1.55~K because of pressure increase in the JT stage. The
calculated shift of 20~Hz was taken into account in the data analysis.

(2) Resonator deformation due to gravity: The sensitivity to tilt
was measured to be $2.5\times10^{-16}/\mu$rad and $1.5\times10^{-17}/\mu$rad
for tilt around the longitudinal and transverse axis, respectively.
A tilt control system actuated two of the legs supporting the whole
cryostat and reduced the tilt instability to a level below $0.5\,\mu$rad
for integration times between 100~s and $10^{4}$~s.

The time variation of the local gravitational acceleration (tides,
etc.) has a negligible effect on the resonator. 

(3) Laser power: the laser power incident onto the resonator during
line scan interrogation was 30~$\mu$W or less, and was not actively
stabilized. The power actually coupled into the resonator was 1.5\%
of the incident power. The power absorbed in the mirror substrate
as the laser wave traverses it before entering the resonator and the
power dissipated inside it could be detected via the concomitant resonator
temperature changes. However, the corresponding thermal expansion
is negligible. No effect on the resonator frequency could be detected
directly.

(4) Vibrations: They are caused by the periodic (0.7~s) pulsing of
the PTC and were characterized at room temperature by motion sensors
attached to a plate close to the plate supporting the resonator).
All spatial components of the acceleration have complex time evolutions
(Fourier frequencies up to a few 100~Hz), and r.m.s. values of $(1-8)\times10^{-3}\, g$
within a sensor band-width of 200~Hz. The accelerations cause resonator
deformations that lead to fractional frequency shifts on the order
of $10^{-12}$, periodic in time. We also performed an interferometric
measurement of the periodic axial displacement of a second, identical
silicon resonator inside the cryostat. The amplitude is approximately
$10~\mu$m. 

\begin{figure}[t]
\begin{minipage}[t]{1\columnwidth}%
\noindent \begin{center}
\includegraphics[width=1\columnwidth]{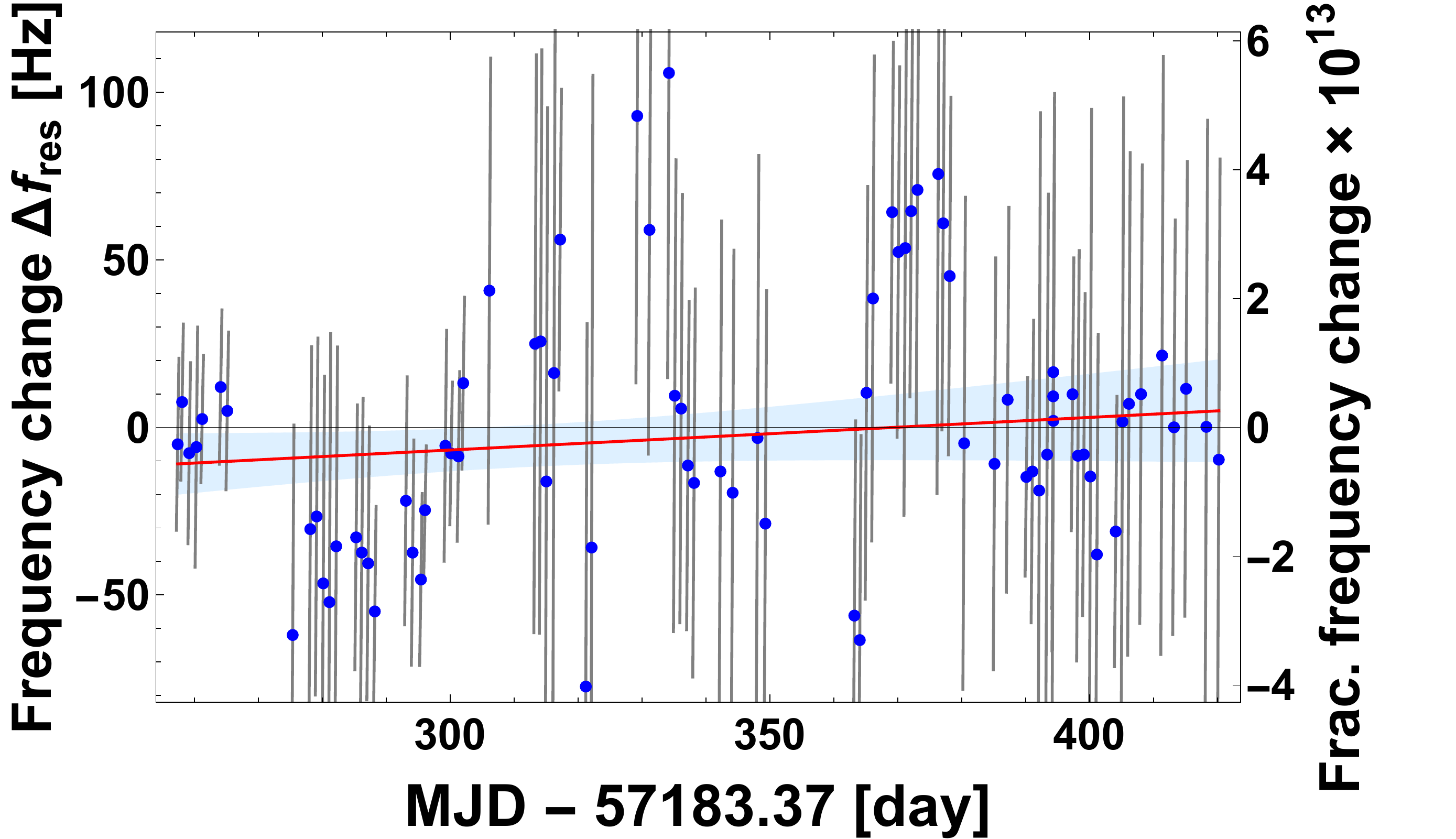}
\par\end{center}%
\end{minipage}

\protect\caption{\label{fig:Si1 resonator frequency zooms} (Color online). Resonator
optical frequency variation $\Delta f_{res}$, corrected for maser
drift, during time interval VII, defined in Fig.~\ref{fig:Si resonator frequency}.
Each data point (blue) is the average frequency value obtained during
a measurement interval $i$. The bars indicate the range $\pm$ twice
the standard deviation of each data set $i$. Red line: time-linear
fit, exhibiting a drift rate $D_{res-atomic}=5.9\times10^{-21}/$s;
blue shaded area: $2\sigma$ uncertainty range of the time-linear
fit.  Zero ordinate value is defined as the mean of the data points. }
\end{figure}

(5) Resonator interrogation by the frequency-scan technique: This
technique (Fig.~\ref{fig:Comb Lock-1} and \cite{SupplementalMaterial-Cryoresonator2016-for-arxiv-tex-version})
leads to an Allan deviation of the line center frequency of $2\times10^{-14}$
at 1000~s integration time. During each frequency scan across the
resonator mode, the signal recorded by detector PD1 is modulated (25\%
fractionally) in time in synchronism with the PTC pulsing, probably
due to variations in coupling efficiency caused by the pulsing. The
modulation is complex, with pulse-to-pulse variations. These signal
disturbances lead to variations of the line centers frequencies fitted
from individual scans. These variations are of the order $3\times10^{-13}$
(2$\sigma$ of the data) (gray bars in Fig.~\ref{fig:Si1 resonator frequency zooms}).

(6) Long-term effects of laser light. The laser intensity circulating
in the resonator might cause photochemical or structural changes in
the mirror coatings, with consequent resonator frequency change. We
did not keep any laser frequency resonant for extended duration, in
order to limit the irradiation of the mirrors \cite{SupplementalMaterial-Cryoresonator2016-for-arxiv-tex-version}. 

(7) Although it is known that maser frequency drift magnitude can
be below $1\times10^{-21}/$s \cite{Matsakis2005}, it is fundamental
to determine the influence of our particular maser on the optical
frequency measurement. The maser was monitored by comparison with
a 1-pulse-per-second signal delivered by GPS, which is derived from
the international atomic time scale (TAI), defined by the cesium hyperfine
transition. During the period day 225 - 264 the mean fractional drift
was $8.2\times10^{-21}/$s. During the period day 293 - 415, the mean
drift was $D_{maser-GPS}=7.5\times10^{-21}/$s. This level is relevant
in comparison with the drift of the resonator's optical frequency
with respect to the maser, $D_{res-maser}$, and thus must be accounted
for. The error of $D_{maser-GPS}$ is negligible for the present discussion. 

\textit{Measurement of the resonator drift.} Measurements were performed
daily, when possible. In practice, each measurement of $f_{res}$
was taken as a time-average over several minutes, so as to suppress
disturbances occurring over short time scales. The long-term behavior
of the resonator frequency $f_{res}$ is depicted in Fig\@.~\ref{fig:Si resonator frequency}(a).
Frequency jumps $J_{1}$ to $J_{4}$ were caused by large temperature
variations or by unintentional knocks on the cryostat. On day 149
$(J_{5})$, we deliberately knocked on the cryostat's outer vacuum
chamber, and observed 56~kHz frequency change. The frequency jump
after $J_{6}$ was probably caused by a power shut-down \cite{SupplementalMaterial-Cryoresonator2016-for-arxiv-tex-version}.
To compensate for these jumps we shifted all frequency values after
each jump by an amount equal to the difference between the last measurement
before and first measurement after each jump. In Fig\@.~\ref{fig:Si resonator frequency}(b)
we display the resulting time series of the corrected frequency $f_{res}$,
the offsets for {[}$J_{1}$,...,$J_{6}${]} being $[9.1,7.8,2.5,1.4,56.1,0.4]\pm0.1$~kHz.

We find that within 2~weeks after each disturbance $J_{1}$ to $J_{6}$,
the drift has dropped back to nominally zero. The data intervals III
- VI are intervals between frequency jumps during which the resonator
remained undisturbed. During these intervals, the drift rates were
compatible with zero with upper limits $4\times10^{-19}$/s.  Interval
VII is the period of longest undisturbed duration, 163~days. Starting
with this interval, the frequency-scan interrogation technique was
used. The $f_{res}$ data thus obtained is shown in Fig.~\ref{fig:Si1 resonator frequency zooms}.
The variations of $\Delta f_{res}$ are mostly due to systematic effects
(1) - (6); a possible additional contribution are long-term (time
scale of days) maser frequency fluctuations, which we cannot independently
identify via GPS, due to latter's low stability. 

Given the long overall measurement duration of 163~days we can assume
that the variations are approximately randomly distributed. A time-linear
fit of the optical frequency data $f_{res}$ yields $D_{res-maser}=(-1.6\pm3.8)\times10^{-21}$/s
(1~$\sigma$).  We then obtain the drift of the silicon resonator
frequency with respect to atomic time as $D_{res-atomic}=D_{res-maser}+D_{maser-GPS}=$
$(5.9\pm3.8)\times10^{-21}$/s. The systematic effects disturbing
the laser frequency scan data and the finite overall measurement time
span determine the uncertainty of $D_{res-atomic}$.
\textit{}

\textit{Interpretation.} We interpret the zero drift in three ways:
(a) as a test of LPI \cite{Turneaure1983}, (b) as a test for effects
related to the expansion of the universe, and (c) as a test of the
existence of space-time fluctuations. Local Lorentz Invariance is
assumed to hold.

(a) One test of LPI is a null clock redshift test that tests for deviations
from the clock-type-independence of the gravitational time dilation.
It consists in measuring the ratio of the frequencies of two dissimilar
clocks $C1,\, C2,$ co-located at ${\bf r}$, as they are transported
through a region of varying gravitational potential $U({\bf r})$.
The change of the frequency ratio is written as

\noindent 
\begin{equation}
(f_{C1}/f_{C2})({\bf r})=(f_{C1}/f_{C2})_{0}(1+(\xi_{C1}-\xi_{C2})\Delta U({\bf r})/c^{2})\,,\label{eq:frequency ratio}
\end{equation}
where $\xi_{C1},\,\xi_{C2}$ are the gravitational coupling constants
of the two clock types, $(f_{C1}/f_{C2})_{0}$ is the frequency ratio
measured for the arbitrary reference value $U_{0}$, and $\Delta U({\bf r})=U({\bf r})-U_{0}$.
If General Relativity holds, $\xi=1$, independent of the type of
clock. For two co-located clocks on Earth, $U({\bf r})=U_{solar}({\bf r})$
is time-varying because of rotational and orbital motion. The $\Delta f_{res}$
data of Fig.~\ref{fig:Si1 resonator frequency zooms} essentially
corresponds to $\Delta(f_{res}/f_{atomic})$, the variation of the
ratio of the resonator frequency and the frequency corresponding to
the atomic unit ot time (delivered via GPS). A fit of Eq.~\ref{eq:frequency ratio}
to the data yields $\xi_{res}-\xi_{atomic}=(-2.8\pm1.6)\times10^{-4}$
for the clock coupling to the Sun's gravitational potential, which
is compatible with zero (the error given is the standard error). The
upper bound $|\xi_{res}-\xi_{atomic}|_{2\sigma}<6\times10^{-4}$ is
nearly equal to the result of ref.~\cite{Tobar2010} (accounting
for the result of ref.~\cite{Bauch2002}), where a cryogenic microwave
sapphire oscillator (CSO) was used. However, our work achieved this
result with a measurement duration of 5 months compared to 6 years.

(b) The second interpretation provides a limit for (hypothetical)
linear-in-time effects on measuring rods, caused by the expansion
of the universe, if General Relativity is violated. Ref.~\cite{Kopeikin2015}
discusses that such an effect is absent if General Relativity holds.
To arrive at such a bound we first discuss how bounds of LPI violation
and time-variation of fundamental constants contribute.

We assume that the LPI bound of refs.~\cite{Bauch2002,Tobar2010},
$|\xi_{CSO}-\xi_{atomic}|_{2\sigma}<5.5\times10^{-4}$, holds also
for standing-wave optical resonators, like the one used here. This
implies that during the interval from day 258 to day 420 the mean
resonator drift from a LPI violation is smaller than $|D_{LPI}|_{2\sigma}=1.2\times10^{-20}/$s.
Note that ref.~\cite{Tobar2010} derives the LPI bound from the CSO's
annual frequency modulation amplitude only. The CSO's frequency drift,
$-1.7\times10^{-18}/$s, was removed before data analysis, therefore
the value of $D_{LPI}$ is independent of any expanding-universe effect.
Furthermore, we recall that within conventional physics, $f_{res}/f_{atomic}$
(where $f_{atomic}$ is derived from the Cs hyperfine transition)
can in principle be expressed in terms of the fine-structure constant,
$m_{e}/m_{p}$, nuclear parameters, the number of lattice planes in
the silicon resonator, and numerical coefficients. The current experimental
limits for the drifts of these constants and parameters, derived from
comparisons between atomic clocks \cite{Uzan2011}, lead to a bound
much smaller than our $|D_{res-atomic}|_{2\sigma}$ and $|D_{LPI}|_{2\sigma}$. 

Thus, the results of $D_{res-atomic}$ and $D_{LPI}$ can be combined
to set the bound $2.8\times10^{-20}$/s for the magnitude of linear-in-time
drifts of the length of a solid, when measured by clocking the propagation
of electromagnetic waves across its length. This is a factor 82 smaller
than the natural scale $H_{0}$, and thus rules out any effect that
is of first order in $H_{0}$. 

(c) Space-time fluctuations (or ``foam'') is a concept that describes
the possibility that repeated measurements of a particular time interval
or a particular distance do not give a constant result but fluctuate
due to fundamental reasons. The measurement of the resonator frequency
$f_{res}$ performed here ultimately corresponds to a measurement
of a particular distance (the mirror spacing $l_{0}$), in units of
the wavelength of an electromagnetic wave resonant with the cesium
hyperfine transition. Simple models for the noise power spectral density
$S(f)$ ($f$ is the fluctuation frequency) of the fractional length
fluctuations $\Delta l/l_{0}$ have been introduced \cite{Ng2003},
leading to flicker frequency noise, $S_{flicker}(f)=\alpha/f$ or
random-walk frequency noise, $S_{rw}(f)=\beta/f^{2}$. Experiments
are required to place upper limits on $\alpha$ and $\beta$.

We compare our time series of $f_{res}$ with simulated time series
generated from flicker and random walk noise and therefrom deduce
$S_{flicker}(f)<4\times10^{-27}/f$ and $S_{rw}(f)<9\times10^{-33}\,{\rm Hz}/f^{2}$.
These limits are weaker than our previous measurements \cite{Schiller2004,Chen2016},
but are here established from data at significantly lower Fourier
frequencies, $f\simeq1\,\mu$Hz.

\textit{Summary and Outlook.} In this work we have demonstrated that
a silicon crystal exhibits an extremely small length drift at cryogenic
temperature. The mean fractional drift $D_{res-atomic}$ measured
during a time span of 5~months of nearly undisturbed operation was
$(5.9\pm3.8)\times10^{-21}$/s. This is the lowest value measured
so far for resonators, to the best of our knowledge. The uncertainty
of the value $D_{res-atomic}$ is due to the systematic effects caused
by the cryostat cooler and by the finite measurement time span. Both
aspects could be improved in the future. The measurement rules out
local consequences of the expansion of the universe which are of
order of the Hubble constant $H_{0}$. In addition, the data provides
the best upper limit for violations of Local Position Invariance for
an optical resonator and for the existence of space-time fluctuations
with frequencies in the $\text{\ensuremath{\mu}}$Hz-range. 

To illustrate the smallness of the drift, we note that the $2\sigma$
upper limit for $D_{res-atomic}$, the positive value $1.4\times10^{-20}$/s,
is consistent with a shortening of the optical path length in the
resonator. If this were caused by deposition of molecules on the mirrors,
the deposition rate would be one molecular layer on each mirror every
$3\times10^{3}$ years, or approximately 30~molecules/s within the
laser beam cross section (1~mm diameter).

 The cryogenic silicon resonator could potentially be used as a local
oscillator with performance beyond that of hydrogen masers and with
autonomous operation. Future progress in cryostat technology may eventually
allow suppressing the vibrations to a sufficiently low level so that
a silicon resonator exhibits sub-$10^{-17}$ fractional frequency
instability, on all time scales, in addition to negligible drift.
Until this potential is realized, another suggested implementation
consists of a cryogenic silicon resonator combined with an improved
version of the ULE-resonator-stabilized laser used here \cite{SupplementalMaterial-Cryoresonator2016-for-arxiv-tex-version}.
State-of-the-art ULE resonators are in principle capable of providing
frequency instability at $1\times10^{-16}$ level for integration
times of up to 1000~s (see, e.g. \cite{Jiang2011,Nichloson2012,Haefner2015}),
but this requires eliminating their drift, of typical magnitude $1\times10^{-16}$/s.
The drift would be determined by periodic comparison with the silicon
resonator and would be compensated by a feed-forward scheme. Our work
suggests that the proposed solution will be robust and practical,
since  we showed uninterrupted operation of the resonator for 12~months
without serious problems.     
\begin{acknowledgments}
We thank I.~Ernsting, Q.-F.~Chen, U.~Rosowski, D.~Iwaschko,
P.~Dutkiewich, and R.~Gusek for valuable contributions, and M.~Schioppo
for comments. This work has been funded by ESA project no.~4000103508/11/D/JR
and by DFG project Schi~431/21-1. 

E.W. and S.S. contributed equally to this work.\phantom{\cite{Braxmaier2000}}Corresponding
author: \textit{step.schiller@hhu.de}\end{acknowledgments}

\end{document}